\documentstyle[12pt, A4]{article}
\thicklines                     
\begin{document}
\begin{titlepage}
\begin{centering}
\vspace{2cm}     
{\Large\bf Generating functionals of correlation functions }\\
\vspace{0.5cm}
{\Large\bf of $p$-form currents in AdS/CFT correspondence}\\
\vspace{2cm}
W.~S.~l'Yi\footnote{E-mail address: {\tt wslyi@cbucc.chungbuk.ac.kr}}
\vspace{0.5cm}

{\em Department of Physics}\\
{\em Chungbuk National University}\\
{\em Cheongju, Chungbuk 361-763, Korea}\\
\vspace{1cm}
\begin{abstract}
\noindent                    
The generating functional of correlation functions of the currents 
corresponding to general massless $p$-form potential is calculated 
in $AdS/CFT$ correspondence of Maldacena.  For this we construct 
the boundary-to-bulk Green's functions of $p$-form potentials. The proportional
constant of the current-current correlation function, which is related to the 
central charge of the operator product expansion, is shown to be 
$c={d-p\over 2\pi^{d\over 2}}{\Gamma (d-p) \over \Gamma ({d\over 2}-p)}.$
The result agrees with the known cases such as $p=1$ or 2.
\end{abstract}
\end{centering} 
\vfill
\noindent HEP-CNU-9810\\
hep-th/9809132\\
Sept.~1998
\end{titlepage}

\setcounter{footnote}{0}
\section{Introduction}
Ever since the introduction of Maldacena's proposal\cite{Maldacena} that
the classical supergravity action on Anti-de Sitter space ($AdS$) can be used 
to determine the current-current correlation function of the dual conformal 
field theory
($CFT$) on the boundary of $AdS$\cite{Gubser, Witten_holography},
various successful applications are 
reported in the direction of confirming it\cite{applications,Horowitz,black_hole}.

Two point correlators for the case of scalar 
fields\cite{Gubser, Witten_holography, scalar_theory,Ya_1,Ya_2}, 
three\cite{Freedman1} and 
four-point correlators\cite{Tseytlin,Freedman2} are investigated.
For massless vectors it is rather simple when one benefits from the gauge 
transformations to eliminate the ${\cal A}_0$ component of 
$AdS$ vector potential\cite{Witten_holography}.
But for massive vector fields there is no such transformations.  In fact 
the determination of ${\cal A}_0$ component of a vector potential is an  
essential element of holographic projection of massive vector 
fields\cite{MV,lyi}.
Spinors\cite{Leigh,Henningson,MV}, gravitinos\cite{Corley,Volovich}, 
and gravitons\cite{Arutyunov_1,Tseytlin}
are also investigated.   
The next natural step is the consideration 
of $p$-form potential.  Since supergravity actions and brane theories contain
various $p$-form potentials, it is rather one of the imperative for the full 
understanding of $AdS/CFT$ correspondence.  

In this paper the generating functional of current-current correlation
functions of general $p$-form potentials are constructed using 
the $AdS/CFT$ correspondence.
We follow Witten's intuitive methodology\cite{Witten_holography}
to construct the boundary-to-bulk Green's function.  But rather than 
delving into the $p$-form potential we begin with the 2-form case.  
Gauge transformation of 2-form potential is used to elemenate unnessesary
components of the potential.  
The boundary-to-bulk Green's function which is in fact a 2-form 
is constructed
using the differential form language.  The classical action which is served 
as the generating functional of current-current correlation functions is
determined from this 2-form Green's function.  This is presented in the next 
section.  The generization to the general $p$-form case is rather straight
forward.  This is done in section 3. Appendix is devoted to prove that the 
Witten's method causes no Ward identity problem for massless $p$-form cases
which we consider.   
\section{General consideration of boundary-to-bulk Green's function}
According to the Maldacena's proposal the supergravity partition 
function ${\cal Z}_{AdS_{d+1}}[\phi_a]$ of general fields $\phi_a$ on 
$AdS_{d+1}$ can be used as the generating functional 
$\langle \exp \int d^dx \varphi_i({\bf x}) {\cal O}_i({\bf x}) \rangle_{CFT}$ 
of the operator product expansions of sources 
${\cal O}_{i}({\bf x})$ on 
the boundary of $AdS.$
The relation between $\phi_a$ and $\varphi_i$ is that when one goes
to the boundary of $AdS,$ $\phi_a$ reduces to $\varphi_i.$
When the classical approximation is employed to the supergravity partition
function,
Maldacena's proposal implies that the classical action $I[\phi_a]$ of the 
bulk field $\phi_a,$
when written in terms of boundary
field $\varphi_i,$ serves as the generating functional of the OPE.
This means that it is essential for the calculation of OPE of the sources 
${\cal O}_{i}({\bf x})$ to determind the boundary-to-bulk Green's
function. 

For our purpose it is sufficient to employ the Euclidean $AdS$ which is
characterized by the Lobachevsky space
${\bf R}^{1+d}_+ = \{(x_0,{\bf x})\in {\bf R}^{1+d}\; | \;x_0>0\}$
with the Poincare metric of the following form,
\begin{equation}
ds^2={1\over x_0^2}\left[ (dx^0)^2 + (d{\bf x})^2\right]. \label{metric}
\end{equation}  
The $x_0\to\infty$ point and the $x_0\to 0$ region consist the boundary of 
$AdS.$
To simplify the notation we sometimes confuse $x^0$ with $x_0$ unless 
it is stated explicitly. Roman charactors such as $i$ and $j$ are used 
to denote the indices $1,...,d,$ and Greek characters such as $\mu$ and $\nu$
are preserved to denote whole indices $0,1,...,d.$

It is well known that to have correct correlation functions which preserve 
the Ward identity one should be careful to have all fields to approach the 
boundary of AdS uniformly.  For this an imaginative $x_0=\epsilon$ boundary 
is first used to compute the classical action, and
then one take the $\epsilon\to0$ limit.  In \cite{lyi} this procedure
in coordinate-space is analysed.  Using this algorism one may prove that for 
massless $p$-form field we do not need to worry about the $\epsilon$-boundary 
procedure.  For more details, see appendix.

\section{The generating function of OPE of 2-form currents}

Consider a 2-form potential ${\cal A}$ in $AdS.$
The free action is given by
\begin{equation}
I={1\over 2}\int_{AdS_{d+1}} {\cal F}\wedge{}^{*}{\cal F},
\end{equation}
where ${\cal F} = d{\cal A}$ is the field strength 3-form.
The classical equation of motion of ${\cal A}$ obtained from this action is 
\begin{equation}
d^{*}d{\cal A} = 0.
\end{equation}
To calculate the current-current correlation function in $AdS/CFT$ correspondence
we need to know the boundary-to-bulk 2-form Green's function,
\begin{equation} 
K(x_{bk},x_{bd})={1\over 2}K_{\mu\nu}(x_{bk},x_{bd})dx_{bk}^\mu \wedge 
dx_{bk}^\nu.
\end{equation}
Here $x_{bk}$ is a point in $AdS,$  and $x_{bd}$ is a point on the boundary
of it.  $K$ satisfies the same equation which 
${\cal A}$ satisties, and its component becomes the boundary-space 
delta function as $x_{bk}$ approaches $x_{bd}.$
Using the gauge transformation,
\begin{equation}
{\cal A} \to {\cal A} + d\Lambda,
\end{equation}
of a 2-form potential we may assume that all $K_{0i},$ $i=1,...d,$ vanish. 
For simplicity we take as $x_{bd}$ the boundary point at $x_0\to\infty.$  
The boundary of $AdS$ and the Poincare metric are both invariant
under the following translation,
\begin{equation}
{\bf x} \to {\bf x} - {\bf x}',\label{translation}
\end{equation}                    
where ${\bf x}'$ is a constant vector.  This allows us to assume that 
$x_{bk}$ is $(x_0,0,...,0)$\cite{Witten_holography}.
The general form of $K$ under these restrictions is
\begin{equation}
K_\infty=K_\infty(x_0)dx^i \wedge dx^j,
\end{equation}
where $i<j.$ The infinity symbol in this equation denotes the fact the 
the boundary point of $K$ is chosen to be $x_0 \to \infty.$

The equation which is satisfied by $K_\infty$ is 
\begin{equation}
d^{*}dK_\infty = 0.
\end{equation}
To solve this we compute $dK_\infty,$
\begin{equation}           
dK_\infty=\partial_0 K_\infty(x_0) dx^0 dx^i dx^j,
\end{equation}
and use the relation,
\begin{equation}
{}^*(dx^0 dx^i dx^j) = (-)^{i+j+1} x_0^{-(d-5)} 
\widehat{dx^0} dx^1...\widehat{dx^i}... \widehat{dx^j}...dx^d,
\end{equation}
where $\widehat{dx^{0}},$ etc., means that $dx^{0}$ should be omitted 
in the wedge product.  The result is 
\begin{equation}
\partial_0 \left(x_0^{-(d-5)} \partial_0 K_\infty(x_0)\right) = 0,
\end{equation}
whose solution is 
\begin{equation}
K_\infty=\alpha x_0^{d-4}dx^i dx^j.
\end{equation}
Here $\alpha$ is an unknown constant. 

For the case $d > 4,$ it is clear that as $x_0$ goes to 0, $K_\infty (x_0)$ 
vanishes.  When $x_0$ goes to $\infty,$ $K_\infty (x_0)$ diverges.  
This means that $K_\infty$ looks like a boundary-to-bulk Green's function.  
Before proving this fact explicitly we benefit from the following
gauge transformation, 
\begin{equation}
K_\infty \to K_\infty + d(\beta {x_0}^{d-4}x^i dx^j - \gamma {x_0}^{d-4}x^j dx^i).
\end{equation}                                                     
The transformed $K_\infty$ is given by
\begin{equation}
K_\infty=(\alpha +\beta +\gamma) x_0^{d-4} dx^i dx^j 
    + (d-4)\beta  x_0^{d-5}x^i dx^0 dx^j
    - (d-4)\gamma x_0^{d-5}x^j dx^0 dx^i.
\end{equation}
By choosing the coefficients which satisfy
\begin{equation}
\alpha + \beta + \gamma = 1, \;\;\;
\beta = \gamma = -{1\over d -4},
\end{equation}
$K_\infty$ reduces to
\begin{equation}
K_\infty = x_0^{d-4} dx^i dx^j  - x_0^{d-5} x^i dx^0 dx^j
   + x_0^{d-5} x^j dx^0 dx^i.
\end{equation}
It is still possible to rescale $K_\infty$ by a constant $c$ without losing
the desired property.

To prove that $K_\infty$ in fact generates the correct 
boundary-to-bulk Green's function
we use the following isometry of (\ref{metric}) which is discussed in detail in
\cite{Freedman1},
\begin{equation}
x^\mu \to {x^\mu \over x_0^2 + {\bf x}^2}.\label{isomorphism}
\end{equation}  
It is important to note that under this isometry the boundary point 
corresponding to $x_0\to\infty$ transforms to the boundary point 
$(x_0,{\bf x})\to 0.$
This isometry transforms $K_\infty$ to
\begin{equation}
K_0={x_0^{d-4} \over (x_0^2 +  {\bf x}^2)^{d-2} } dx^i dx^j
    - {x_0^{d-5} x^i \over (x_0^2 +  {\bf x}^2)^{d-2} } dx^0 dx^j
    + {x_0^{d-5} x^j \over (x_0^2 +  {\bf x}^2)^{d-2} } dx^0 dx^i,\label{K_0}
\end{equation}
where the subscript ``0'' in $K_0$ means that the boundary point of $K_0$ 
corresponds to $(x_0,{\bf x})\to 0.$
To get the Green's function corresponding to a 
general boundary point
$(0,{\bf x}')$ we use the translational symmetry (\ref{translation}).
This gives the final Green's function,
\begin{equation}
cK={x_0^{d-4} \over (x_0^2 +  |{\bf x}-{\bf x}'|^2)^{d-2} } dx^i dx^j
    - {x_0^{d-5} x^i \over (x_0^2 +  |{\bf x}-{\bf x}'|^2)^{d-2} } dx^0 dx^j
    + {x_0^{d-5} x^j \over (x_0^2 +  |{\bf x}-{\bf x}'|^2)^{d-2} } dx^0 dx^i,
 \label{cK}
\end{equation}
where $c$ is a normalization constant.
On the other hand, it is not difficult to show that
\begin{equation}
\lim_{x_0\to 0} {x_0^{d-2\nu} \over (x_0^2 + |{\bf x}-{\bf x}'|^2)^{d-\nu} }
=c_{d,\nu}\delta^{(d)}({\bf x}-{\bf x}'),
\end{equation}
where, 
\begin{equation}
c_{d,\nu}=\pi^{d\over 2}{\Gamma({d\over 2}-\nu) \over \Gamma(d-\nu)}.
\end{equation}
This means that by choosing the constant $c$ of (\ref{cK}) which is equal  
to $c_{d,2},$ $K$ of (\ref{cK}) is in fact the desired boundary-to-bulk 
Green's function.
                                                                       
Now let $A = {1\over 2}A_{ij}({\bf x}) dx^i dx^j$ be a 2-form potential
on the boundary of $AdS.$   Using (\ref{cK}) we have the following 
2-form bulk potential,
\begin{eqnarray}
c_{d,2}\, {\cal A}(x_0, {\bf x}) &=& \left[x_0^{d-4}
 \int d^d x' { A_{ij}({\bf x}') \over (x_0^2 + |{\bf x} - {\bf x}'|^2)^{d-2} } 
  \right] {1\over 2} dx^i dx^j\\
&&-2 \left[ x_0^{d-5}
 \int d^d x' { (x^i-x'^i)A_{ij}({\bf x}') 
       \over (x_0^2 + |{\bf x} - {\bf x}'|^2)^{d-2} } 
  \right]{1\over 2}dx^0 dx^j, \nonumber
\end{eqnarray}               
where we used the anti-symmetric property of $A_{ij}.$
The field strength 3-form ${\cal F}$ is given by
\begin{eqnarray}
c_{d,2}\,{\cal F}(x_0, {\bf x}) &=& (d-2)x_0^{d-5}
    \int d^dx' {A_{ij}({\bf x}') \over (x_0^2 + |{\bf x} - {\bf x}'|^2)^{d-2} }
     {1\over 2}dx^0 dx^i dx^j\\
  && - 2(d-2)x_0^{d-3}
    \int d^dx' {A_{ij}({\bf x}') \over (x_0^2 + |{\bf x} - {\bf x}'|^2)^{d-1} }
     {1\over 2}dx^0 dx^i dx^j \nonumber \\
  && - 4(d-2)x_0^{d-5}
    \int d^dx' {(x^j-x'^j) (x^k-x'^k) A_{ij}({\bf x}')  \over 
               (x_0^2 + |{\bf x} - {\bf x}'|^2)^{d-1} }{1\over 2}dx^0 dx^i dx^k
  + ...\nonumber
\end{eqnarray}
where the abbreviated terms are those which do not contain $dx^0.$  

By partial integration the bulk action becomes
\begin{equation}
I = \lim_{\epsilon \to 0} {1\over 2} \int_{x_0=\epsilon}  
       {\cal A}\wedge {}^*{\cal F}.
\end{equation}
The following relation
\begin{equation}
dx^i dx^j \wedge {}^*(dx^0 dx^l dx^m)
 = x_0^{-(d-5)}\delta^{ij}_{lm}dx^1 ... \,dx^d,\;\;\;
\delta^{ij}_{lm} = \delta^i_l \delta^j_m - \delta^i_m \delta^j_l,
\end{equation}
is useful for the computation of the action.  The final result is given by
\begin{eqnarray}
I[A] &=& c\sum_{i<j} \int d^d x d^d x' 
  { A_{ij}({\bf x}) A_{ij}({\bf x}') \over |{\bf x} - {\bf x}'|^{2(d-2)} } \\
  &&- 2c\sum_{ijk} \int d^d x d^d x'  
   {A_{ij}({\bf x}) A_{ik}({\bf x}')(x_j - x'_j)(x_k - x'_k) 
     \over |{\bf x} - {\bf x}'|^{2(d-1)} },\nonumber
\end{eqnarray}
where
\begin{equation}
c={d-2\over 2\pi^{d\over 2}} {\Gamma (d-2) \over \Gamma ({d\over 2}-2)}.
\end{equation}                                                    
It agrees with the result of \cite{Arutyunov_2}. 
\section{The generating function of OPE of $p-$form currents}
A general $p$-form potential $A$ on the boundary of $AdS$  can be lifted to
a bulk one ${\cal A}$ by using the boundary-to-bulk Green's function $K,$ the 
determination of which is our next step.
Similarly to the 2-form case we assume that
\begin{equation}
K_\infty = K_\infty (x_0) dx^{i_1} ... dx^{i_p},
\end{equation}
where $i_1 < ... <i_p.$
$K_\infty$ satisfies the free-field equation $d^*dK_\infty = 0$ which reduces to
\begin{equation}
\partial_0\left[ x_0^{-(d-2p-1)}\partial_0 K_\infty (x_0) \right] = 0.
\label{equation_of_K}
\end{equation}  
The solution of this equation which vanishes at $x_0\to0$ is
$K_\infty (x_0)=c_0 x_0^{d-2p},$ where $c_0$ 
is a constant. In fact it is true only for
$d>2p.$ But for $d\leq 2p$
one may use dual $p^*$-form such as
$p^*=d-p-1.$  Using this $p^*$ we have the relation $d-2 \geq 2p^*,$ or 
the relation $d > 2p^*$ is satisfied.

To simplify the final calculation we apply the following gauge 
transformation,
\begin{eqnarray}
K_\infty &\to& c_0 x_0^{d-2p} dx^{i_1} ... dx^{i_p} \\
&& +d\left[ x_0^{d-2p}\left( c_1  x^{i_1} dx^{i_2} ... dx^{i_p} 
  -c_2 dx^{i_1} \cdot x^{i_2}\cdot dx^{i_3}... dx^{i_p} ... \nonumber
 +(-)^{p+1} c_p dx^{i_1} ...  dx^{i_{p-1}}\cdot x^{i_p}\right) \right].
\end{eqnarray}
The transformed $K_\infty$ is given by
\begin{eqnarray}
K_\infty &=& (c_0 +c_1 + \cdots + c_p) x_0^{d-2p} dx^{i_1} ... dx^{i_p} \\
 &&+ c_1(d-2p) x_0^{d-2p-1} x^{i_1} dx^0 \widehat{dx^{i_1}} dx^{i_2} ... dx^{i_p}
    \nonumber\\
 &&- c_2(d-2p) x_0^{d-2p-1} x^{i_2} dx^0 dx^{i_1} \widehat{dx^{i_2}}  ... dx^{i_p}
    \nonumber\\
 &&... \nonumber \\   
 &&+(-)^{p+1} c_p(d-2p) x_0^{d-2p-1} x^{i_p} dx^0 dx^{i_1}...dx^{i_{p-1}}
    \nonumber
  \widehat{dx^{i_p}},                                                     
\end{eqnarray}
where $\widehat{dx^{i_k}},$ $k=1,...,p,$ means that $dx^{i_k}$ should be omitted 
in the wedge product.  When one choose constants such as
\begin{equation}
c_1=c_2=\cdots=c_p = {-1\over d-2p},\;\;\; c_0 = {d-p\over d-2p},
\end{equation}                                                  
$K_\infty$ simplifies to 
\begin{equation}
K_\infty=x_0^{d-2p-1}\sum_{\mu=0}^p (-)^\mu  x^{i_\mu} dx^{i_0}...
         \widehat{dx^{i_\mu}}... dx^{i_p}.
\end{equation}                                                        
Here we use the notation $x^{i_0}=x^0.$  
Under the isomorphism (\ref{isomorphism}) $K_\infty$ transforms to 
\begin{eqnarray}
c_{d,p}K_0 &=& x_0^{d-2p-1} \sum_{\mu=0}^p (-)^\mu 
   {x^{i_\mu}\over (x_0^2 + {\bf x}^2)^{d-p} }
   dx^{i_0}...\widehat{dx^{i_\mu}}... dx^{i_p} \\
&=& {x_0^{d-2p} \over (x_0^2 + {\bf x}^2)^{d-p}} dx^{i_1} ...dx^{i_p}\nonumber\\
 && + x_0^{d-2p-1} \sum_{k=1}^p (-)^k 
   {x^{i_k} \over (x_0^2 + {\bf x}^2)^{d-p} }
  dx^{0}dx^{i_1}...\widehat{dx^{i_k}}...dx^{i_p}, \nonumber
\end{eqnarray}                                   
where the normalization constant $c_{d,p}$ is multiplied to get the correct $K.$
The final step for the construction of the boundary-to-bulk Green's function 
is to use the translation (\ref{translation}).

Consider a $p$-form potential on the boundary of $AdS,$ 
\begin{equation}
A = {1\over p!}\sum A_{i_1 ... i_p}dx^{i_1} ... dx^{i_p}.
\end{equation}
The bulk potential ${\cal A}$ which is lifted from this is
\begin{eqnarray}
c_{d,p}\,{\cal A}(x_0, {\bf x}) &=& \sum_{i_1...i_p}x_0^{d-2p}\int d^d x'
{A_{i_1 ... i_p}({\bf x}') \over (x_0^2 + |{\bf x} - {\bf x}'|^2)^{d-p} }
    {1\over p!} dx^{i_1} ... dx^{i_p} \\
&+& \sum_{i_1...i_p}\sum_{k=1}^p (-)^k x_0^{d-2p-1}\int d^d x'
{(x_{i_k} - x'_{i_k}) A_{i_1 ... i_p}({\bf x}') \over (x_0^2 + |{\bf x} - {\bf x}'|^2)^{d-p} }
    {1\over p!} dx^0 dx^{i_1} ... \widehat{dx^{i_k}} ... dx^{i_p}.\nonumber
\end{eqnarray} 
We confuse notation $x_{i_k}$ with $x^{i_k}$ for convenience.
Using the anti-symmetric property of $A_{i_1 ... i_p}$ it can be further simplified
to
\begin{eqnarray}
c_{d,p}\,{\cal A}(x_0, {\bf x}) &=& \sum_{i_1...i_p}x_0^{d-2p}\int d^d x'
{A_{i_1 ... i_p}({\bf x}') \over (x_0^2 + |{\bf x} - {\bf x}'|^2)^{d-p} }
   {1\over p!} dx^{i_1} ... dx^{i_p} \\
&-& p \sum_{i_1...i_p} x_0^{d-2p-1}\int d^d x'
{(x_{i_1} - x'_{i_1}) A_{i_1 ... i_p}({\bf x}') 
  \over (x_0^2 + |{\bf x} - {\bf x}'|^2)^{d-p} }
    {1\over p!} dx^0 dx^{i_2} ... dx^{i_p}.\nonumber   \label{gereral_A}
\end{eqnarray}
The field strength $(p+1)$-form ${\cal F} = d{\cal A}$ is given by
\begin{eqnarray}
c_{d,p}\,{\cal F} &=& (d-p) x_0^{d-2p-1}\sum_{i_1...i_p}
  \int d^d x'
{A_{i_1 ... i_p}({\bf x}') \over (x_0^2 + |{\bf x} - {\bf x}'|^2)^{d-p} }
   {1\over p!} dx^{0}dx^{i_1} ... dx^{i_p} \\
&&-2(d-p) x_0^{d-2p+1}\sum_{i_1...i_p}\int d^d x'
{A_{i_1 ... i_p}({\bf x}') \over (x_0^2 + |{\bf x} - {\bf x}'|^2)^{d-p+1} }
   {1\over p!} dx^{0}dx^{i_1} ... dx^{i_p} \nonumber \\     
&&-2p(d-p) x_0^{d-2p-1} \sum_{iji_2...i_p} \int d^d x'
{(x_{i} - x'_{i}) (x_{j} - x'_{j}) A_{i i_2 ... i_p}({\bf x}') 
   \over (x_0^2 + |{\bf x} - {\bf x}'|^2)^{d-p+1} }
    {1\over p!} dx^0 dx^j dx^{i_2} ... dx^{i_p}.\nonumber\\
&& + {\cal F}',   \nonumber  \label{gereral_F}
\end{eqnarray}
where ${\cal F}'$ are the terms which do not have $dx^0.$

The action, using the equation of motion, is given by
\begin{equation}
I={1\over 2}\int_{AdS_{b+1}}  {\cal F}\wedge{}^{*}{\cal F}
 ={1\over 2}\lim_{\epsilon\to 0}\int_{x^0=\epsilon}  
{\cal A}\wedge{}^{*}{\cal F}.
\end{equation}
To compute this we make use of 
\begin{equation}                                     
dx^{i_1}... dx^{i_p} \wedge {}^*( dx^0 dx^{j_1}...\, dx^{j_p}) 
 = x_0^{-(d-2p-1)}\delta^{i_1...i_p}_{j_1...j_p} dx^1 ...\, dx^d,
\end{equation}                                                        
where $\delta^{i_1...i_p}_{j_1...j_p}$ denotes the sign of the corresponding 
permutation.

Using (\ref{gereral_A}) and (\ref{gereral_F}) this can be simplified to give
\begin{eqnarray}
I&=& c \sum_{i_1<...<i_p} \int d^d{\bf x}d^d{\bf x}'
{A_{i_1...i_p}({\bf x})A_{i_1...i_p}({\bf x}') 
     \over |{\bf x} - {\bf x}'|^{2(d-p)}}\\
 && -2c \sum_{ij}\sum_{i_2<...<i_p}
   \int d^d{\bf x}d^d{\bf x}' {(x_i-x'_i)(x_j-x'_j)
    A_{ii_2...i_p}({\bf x})A_{ji_2...i_p}({\bf x}') 
     \over |{\bf x} - {\bf x}'|^{2(d-p+1)}},  \nonumber
\end{eqnarray}                                                
where
\begin{equation}
c={d-p\over 2\pi^{d\over 2}} {\Gamma (d-p) \over \Gamma ({d\over 2}-p)}.
\end{equation}
%
\section*{Acknowledgement}
The author is grateful to G.~Arutyunov for introducing the related works
\cite{Arutyunov_1}, \cite{Arutyunov_2}.
This work is supported in part by NON DIRECTED RESEARCH FUND, 
Korea Research Foundation, 1997, and in part by the Basic Science Research
Institute Program of the Ministry of Education, Korea, BSRI-97-2436.
\section*{Appendix}
\renewcommand{\theequation}{A.\arabic{equation}}
\setcounter{equation}{0}      
In this appendix we follow the notation of \cite{lyi}.
Suppose that $\phi(x_0,{\bf x})$ is a field in $AdS$ which has following 
asymptotic behaviour
\begin{equation}
\phi\to x_0^{-\lambda}\hskip.2cm {\textstyle\rm as\ }x_0\to 0.
\end{equation}
Then one defines a holographically projected field $\phi_h({\bf x})$
by
\begin{equation}
\phi_h({\bf x})=\lim_{\epsilon\to0}\epsilon^\lambda\phi(\epsilon,{\bf x}).
\end{equation}
This is exactly the definition which Witten employed in his 
paper\cite{Witten_holography}.
This boundary field is related to the bulk field $\phi(x_0,{\bf x})$ by
\begin{equation}
\phi(x_0,{\bf x}) = \int d^dx'\; K(x_0, {\bf x} - {\bf x}') \phi_h({\bf x}').
\end{equation}
Here $K(x_0, {\bf x} - {\bf x}')$ is a boundary-to-bulk Green's function
which obeys the same equation which $\phi(x_0,{\bf x})$ satisfies, 
and has following boundary condition
\begin{equation}
\lim_{x_0\to 0}K (x_0, {\bf x}-{\bf x}') 
        =\delta^{(d)} ({\bf x}-{\bf x}').
\end{equation}
This Green's function is clearly different from the Green's function
$G_\epsilon (x_0, {\bf x}-{\bf x}')$ defined to relate the original $AdS$ field
$\phi(x_0,{\bf x})$ and $\phi_\epsilon({\bf x}),$ where
\begin{equation}
\phi_\epsilon({\bf x}) = \phi(\epsilon,{\bf x}).
\end{equation}

Even though Witten's procedure is quite practical it may give wrong results
when one considers massive cases.  In fact one should use $G_\epsilon$ rather than
$K$ in order to preserve the Ward identity.
But for massless case there is no such problem.  It comes from the fact that for
massless $p$-form field, $\lambda=0.$  In this case $AdS$ field at the boundary 
$\phi(0,{\bf x})$ 
is exactly equal to $\phi_h({\bf x}),$ and two Green's functions
$G_\epsilon (x_0, {\bf x}-{\bf x}')$ and $K(x_0+\epsilon, {\bf x}-{\bf x}')$ are equal.
It thus causes no problem at all. 

Now we prove that for $p$-form, $\lambda=0.$  
Suppose $\phi$ is a $p$-form with the equation of motion $d^*d\phi=0.$  
The asymptotic
form of this equation, which is exactly the same as (\ref{equation_of_K}) which
the Green's function satisfies, is given by
\begin{equation}
\partial_0\left[ x_0^{-(d-2p-1)}\partial_0 \phi_\infty (x_0) \right] = 0,
\end{equation}  
where $\phi_\infty (x_0)$ is the asymptotic form of independent component of
$\phi.$ The general solution of this equation is
\begin{equation}
\phi_\infty (x_0) = c_1 + c_2 x_0^{d-2p}.
\end{equation}
It is already proven that we may assume that $d>2p$ without loss of any 
generality. This shows that for $p$-form we do not need to introduce
a rescaling factor such as $x_0^\lambda.$  In other word, $\lambda=0.$

\vspace{2cm}

\end{document}